\documentclass[a4paper,12pt]{article}
\usepackage{epsfig}% SRC Specials

\newcommand{\be}{\begin{equation}}
\newcommand{\ee}{\end{equation}}
\newcommand{\ba}{\begin{eqnarray}}
\newcommand{\ea}{\end{eqnarray}}
\newcommand{\ban}{\begin{eqnarray*}}
\newcommand{\ean}{\end{eqnarray*}}
\newcommand{\demi}{\frac{1}{2}}

\begin{document}

\title{\Large\bf Quantum mechanical retrocausation? \\Call for nonlocal causal models!}

\author{{\bf Antoine Suarez}\thanks{suarez@leman.ch}\\ Center for Quantum
Philosophy\\ The Institute for Interdisciplinary Studies\\ P.O. Box
304, CH-8044 Zurich, Switzerland}

%\date{ }

\maketitle

\vspace{2cm}
\begin{abstract}
A new possible version of multisimultaneous causality is proposed,
and real experiments allowing us to decide between this view and
quantum mechanical retrocausation are further discussed. The
interest of testing quantum mechanics against as many nonlocal
causal models as possible is stressed.\\

{\em Keywords:} superposition principle, retrocausation,
superluminal nonlocality, multisimultaneous causality.\\

\end{abstract}

\pagebreak

In a recent paper \cite{as98.1} we have proposed an impact series
experiment and argued the quantum mechanical superposition
principle to be at odds with the causality principle. We show in
this paper that the causal model used in \cite{as98.1} is not the
only possible, and discuss further real measurements that may allow
us to decide between quantum mechanical retrocausation and
multisimultaneous causality.\\

Consider again the setup sketched in Fig.1. Photon pairs are
emitted through down-conversion from a source $S$. Photon 1 enters
the left hand side interferometer and impacts on beam-splitter
BS$_{11}$ before being detected in either D$_{1}(+)$ or D$_{1}(-)$,
while photon 2 enters the 2-interferometer series on the right hand
side impacting successively on BS$_{21}$ and BS$_{22}$ before being
detected in either D$_{2}(+)$ or D$_{2}(-)$. Each interferometer
consists in a long arm of length $L$, and a short one of length
$l$. We assume as usual the path difference set to a value which
largely exceeds the coherence length of the photon pair light, but
which is still smaller than the coherence length of the pump laser
light.\\

%%%%%%%%%%%%%%%%%%%%%%%%%%%%%%
\begin{figure}[t]
\centering\epsfig{figure=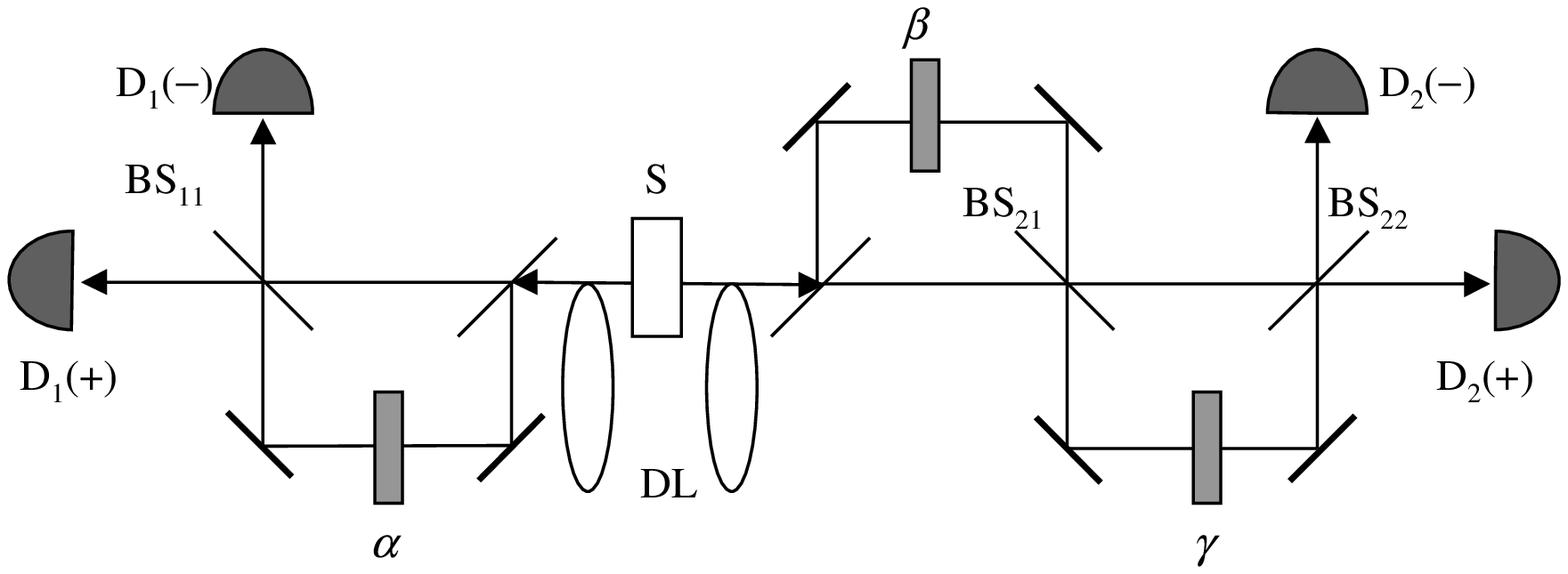,width=120mm}
{\small{\it{\caption{Impact series experiment with photon pairs:
photon 2 impacts successively on beam-splitter BS$_{21}$ and
BS$_{22}$. See text for detailed description.}}}}
\label{fig:BIPfig1}
\end{figure}
%%%%%%%%%%%%%%%%%%%%%%%%%%%%%%

For a pair of photons, eight possible path pairs lead to detection.
We label them as follows: $(l,ll)$; $(L,ll)$; $(l,Ll)$ and so on;
where, e.g., $(l,Ll)$ indicates the path pair in which photon 1 has
taken the short arm, and photon 2 has taken first the long arm,
then the short one.\\

Ordinary Quantum Mechanics assumes indistinguishability to be a
sufficient condition for observing quantum interferences and
entanglement, whereas Relativistic Nonlocality or Multisimultaneity
assumes this condition to be only a necessary one. In any case, as
a first step we must distribute all possible paths in mutually
distinguishable subensembles. The following table gives the four
mutually distinguishable subensembles of the ensemble of all
possible path pairs.\\

\be
\begin{array}{lll}
(l,LL)&:&2L-l\\
(L,LL)\,,\,(l,Ll)\,,\,(l,lL)\,&:&L\\
(l,ll)\,,\,(L,Ll)\,,\,(L,lL)\,&:&l\\
(L,ll)&:&2l-L
\end{array}
\label{eq:paths}
\ee

where the right-hand side of the table indicates the path
difference between the single paths of each photon characterizing
each subensemble of path pairs. From now on, unless stated
otherwise, we consider only those events that are characterized by
path difference $L$, i.e., $(L,LL)\,,\,(l,Ll)\,,\,(l,lL)$.
Experimentally, this is done by appropriate coincidence electronics
\cite{tbg97}. By means of delay lines DL different time orderings
in the laboratory frame can be arranged.\\

The conventional application of the superposition principle yields
the following values for the conventional joint probabilities:

\ba
P^{QM}_{+ +}&=&\frac{1}{12}
\Big[3-2cos(\alpha+\beta)-2cos(\alpha+\gamma)+2cos(\gamma-\beta)\Big]\nonumber\\
P^{QM}_{+ -}&=&\frac{1}{12}
\Big[3-2cos(\alpha+\beta)+2cos(\alpha+\gamma)-2cos(\gamma-\beta)\Big]\nonumber\\
P^{QM}_{- +}&=&\frac{1}{12}
\Big[3+2cos(\alpha+\beta)+2cos(\alpha+\gamma)+2cos(\gamma-\beta)\Big]\nonumber\\
P^{QM}_{- -}&=&\frac{1}{12}
\Big[3+2cos(\alpha+\beta)-2cos(\alpha+\gamma)-2cos(\gamma-\beta)\Big],
\label{eq:qmjp}
\ea

\smallskip

and the corresponding single probabilities for the detections at
side 1 (left-hand side) of the setup:

\ba
P^{QM}_{+\pm}\,\equiv\,P^{QM}_{++}+P^{QM}_{+-}&=&\demi-\frac{1}{3}\cos(\alpha+\beta)\nonumber\\
P^{QM}_{-\pm}\,\equiv\,P^{QM}_{-+}+P^{QM}_{--}&=&\demi+\frac{1}{3}\cos(\alpha+\beta)
\label{eq:qmsp1}
\ea

\smallskip

and at side 2 (right-hand side):

\ba
P^{QM}_{\pm+}\,\equiv\,P^{QM}_{++}+P^{QM}_{-+}&=&\demi+\frac{1}{3}\cos(\beta-\gamma)\nonumber\\
P^{QM}_{\pm-}\,\equiv\,P^{QM}_{+-}+P^{QM}_{--}&=&\demi-\frac{1}{3}\cos(\beta-\gamma)
\label{eq:qmsp2}
\ea

\smallskip

Consider now a multisimultaneous causal model working according to
the following rules:\\

\begin{enumerate}
\item{Half of the pairs traveling by (L,LL) produce outcomes
at BS$_{11}$ and BS$_{22}$ according to superposition with (l,lL),
and half according to superposition with (l,Ll).}
\item{Half of the pairs traveling by (l,lL) produce outcomes
at BS$_{11}$ and BS$_{22}$ according to superposition with (L,LL),
and half according to superposition with (l,Ll).}
\item{Half of the pairs traveling by (l,Ll) produce outcomes
at BS$_{11}$ and BS$_{22}$ according to superposition with (L,LL),
and half according to superposition with (l,lL).}
\end{enumerate}

\smallskip

This model yields the following joint probabilities:

\ba
P^{MC}_{++}&=&\frac{1}{12}
\Big[3-cos(\alpha+\beta)-cos(\alpha+\gamma)+cos(\gamma-\beta)\Big]\nonumber\\
P^{MC}_{+-}&=&\frac{1}{12}
\Big[3-cos(\alpha+\beta)+cos(\alpha+\gamma)-cos(\gamma-\beta)\Big]\nonumber\\
P^{MC}_{-+}&=&\frac{1}{12}
\Big[3+cos(\alpha+\beta)+cos(\alpha+\gamma)+cos(\gamma-\beta)\Big]\nonumber\\
P^{MC}_{--}&=&\frac{1}{12}
\Big[3+cos(\alpha+\beta)-cos(\alpha+\gamma)-cos(\gamma-\beta)\Big],
\label{eq:mcjp}
\ea

\smallskip

and the corresponding single probabilities for the left-hand side:

\ba
P^{MC}_{+\pm}&=&\demi-\frac{1}{6}\cos(\alpha+\beta)\nonumber\\
P^{MC}_{-\pm}&=&\demi+\frac{1}{6}\cos(\alpha+\beta)
\label{eq:mcsp1}
\ea

and for the right-hand side:

\ba
P^{QM}_{\pm+}&=&\demi+\frac{1}{6}\cos(\beta-\gamma)\nonumber\\
P^{QM}_{\pm-}&=&\demi-\frac{1}{6}\cos(\beta-\gamma)
\label{eq:mcsp2}
\ea

\smallskip

Therefore, regarding detections in side 1 the multisimultaneous
causal model proposed in this paper conflicts less with quantum
mechanics than the model proposed in \cite{as98.1}. On the
contrary, whereas the model in \cite{as98.1} did fit with quantum
mechanics for detections at side 2, the model proposed here
conflicts. This clearly supports the retrocausal interpretation of
quantum mechanics for orderings in which the impacts on BS$_{22}$
lie time-like separated after the impacts on BS$_{11}$. Anyway
quantum mechanics seems to exclude any causal explanation.\\

Again, a real experiment can be carried out with the same
arrangement proposed in \cite{as98.1}, i.e., modifying the setup
used in \cite{ptjr94} in order that the photon traveling the long
fiber impacts on a second beam-splitter before it is getting
detected. For the values:

\ba
\alpha+\beta=n\,\pi,\nonumber\\
\beta-\gamma=n\,\pi,
\label{eq:se}
\ea

with $n$ integer, the equations (\ref{eq:qmsp1}) and
(\ref{eq:mcsp1}) yield the predictions:

\ba
E^{QM}=|P^{QM}_{+\pm}-P^{QM}_{-\pm}|=\frac{2}{3},\nonumber\\
E^{MC}=|P^{MC}_{+\pm}-P^{MC}_{-\pm}|=\frac{1}{3},
\label{eq:re1}
\ea

the equations (\ref{eq:qmsp2}) and (\ref{eq:mcsp2}) the
predictions:

\ba
E^{QM}=|P^{QM}_{\pm+}-P^{QM}_{\pm-}|=\frac{2}{3},\nonumber\\
E^{MC}=|P^{MC}_{\pm+}-P^{MC}_{\pm-}|=\frac{1}{3},
\label{eq:re2}
\ea

and the equations (\ref{eq:qmjp}) and (\ref{eq:mcjp}) the joint
probabilities predictions:

\ba
E^{QM}=\sum_{\sigma,\omega}(-\sigma\omega)
P^{QM}_{\sigma\omega}=\frac{2}{3},\nonumber\\
E^{MC}=\sum_{\sigma,\omega}(-\sigma\omega)
P^{MC}_{\sigma\omega}=\frac{1}{3}.\nonumber\\
\label{eq:re3}
\ea

\smallskip

Hence, for settings according to (\ref{eq:se}) the experiment
represented in Fig. 1 allow us again to decide between quantum
mechanics and the multisimultaneous causal model proposed above,
through determining the corresponding experimental quantities from
the four measured coincidence counts $R_{\sigma\omega}$ in the
detectors.\\

Before concluding we would like to point out briefly some of the
possible ramifications the different multisimultaneous causal
models may have. In the type of impact series experiments described
in \cite{as97.3}, the mulsimultaneous causal model proposed in
\cite{as98.1} works according to the quantum mechanical
superposition principle for certain time orderings of the impacts
on the beam-splitters, but permits to arrange 2 {\em non-before}
events with devices at rest in the laboratory frame and,
consequently, conflicts also in this case with quantum mechanics.
On the contrary the multisimultaneous causal model proposed in this
paper works according to the superposition principle only for
experiments without successive {\em non-before} impacts, and is
insensitive to time orderings with beam-splitters at rest. However
it is much more sensitive to time orderings appearing with fast
moving beam-splitters, and may bear plenty of calculation patterns
as those referred to in \cite{as97.1}.\\

In conclusion: Impact series experiments seem to offer a new road
to test Quantum Mechanics against Multisimultaneity with devices at
rest. Since results upholding Quantum Mechanics would strong speak
in favor of retrocausation, impact series experiments could become
with relation to causality what Bell experiments are with relation
to local realism. In the alternative case, the experiments will
allow us to decide which multisimultaneous model fits better the
way Nature behaves. Quantum Mechanics is undoubtedly a pretty well
experimentally confirmed theory, but the questionable nonlocality
of single probabilities revealed by impact series deserves
undoubtedly further testing. To this aim it may be useful to see
whether other suitable nonlocal causal models are still possible,
and to study further the implications of those already proposed.\\

I would like to thank Valerio Scarani (EPFL, Lausanne) and Wolfgang
Tittel (University of Geneva) for stimulating discussions. Support
the L\'eman and Odier Foundations is gratefully acknowledged.

\end{document}